# Observation of nearly identical superconducting transition temperatures in pressurized Weyl semimetal $M$IrTe$_4$ ($M$=Nb and Ta)

Sijin Long[1,4]*, Shu Cai[1]*, Rico Schönemann[2], Priscila F. S. Rosa[2], Luis Balicas[3]，Cheng Huang[1,4]
Jing Guo[1,6], Yazhou Zhou[1], Jinyu Han[1,4], Liqin Zhou[1,4], Yanchun Li[5], Xiaodong Li[5],
Qi Wu[1], Hongming Weng[1,4], Tao Xiang[1,4,6] and Liling Sun[1,4,6]†

[1]*Institute of Physics, National Laboratory for Condensed Matter Physics, Chinese Academy of Sciences, Beijing, 100190, China*
[2]*MPA-MAGLAB, Los Alamos National Laboratory, Los Alamos, New Mexico 87545, USA*
[3]*National High Magnetic Field Laboratory, Tallahassee, Florida 32310, USA*
[4]*University of Chinese Academy of Sciences, Department of Physics, Beijing 100190, China*
[5]*Institute of High Energy Physics, Chinese Academy of Science, Beijing 100049, China*
[6]*Songshan Lake Materials Laboratory, Dongguan, Guangdong 523808, China*

Here we report the observation of pressure-induced superconductivity in type-II Weyl semimetal (WSM) candidate NbIrTe$_4$ and the evolution of its Hall coefficient ($R_H$), magnetoresistance ($MR$), and lattice with increasing pressure to ~57 GPa. These results provide a significant opportunity to investigate the universal high-pressure behavior of ternary WSMs, including the sister compound TaIrTe$_4$ that has been known through our previous studies. We find that the pressure-tuned evolution from the WSM to the superconducting (SC) state in these two compounds exhibit the same trend, *i.e.*, a pressure-induced SC state emerges from the matrix of the non-superconducting WSM state at ~ 27 GPa, and then the WSM state and the SC state coexist up to 40 GPa. Above this pressure, an identical high-pressure behavior, characterized by almost the same value of $R_H$ and $MR$ in its normal state and the same value of $T_c$ in its SC state, appears in both compounds. Our results not only reveal a universal connection between the WSM state and SC state, but also demonstrate that NbIrTe$_4$ and TaIrTe$_4$ can make the same contribution to the normal and SC states that inhabit in the high-pressure phase, although these two compounds have dramatically different topological band structure at ambient pressure.

Weyl semimetals (WSMs) have recently attracted extensive attention in the field of condensed matter physics and material science since Weyl fermions have not been observed so far as a fundamental particle in high energy physics [1-5]. In condensed matter physics, the Weyl semimetal is accepted as a topologically nontrivial metallic phase, where Weyl fermions emerge as quasiparticles from linear crossing bands [1, 6, 7]. Two types of WSMs have been predicted theoretically [1, 8-11] and identified experimentally [3, 12, 13]. The materials of $MX$ ($M$ = Ta, Nb and $X$ = As, P) are classified as type-I WSMs, characterized by a point-like Fermi surface at the Weyl node. While type-II WSMs can be found in the transition-metal dichalcogenide family, which includes the compounds of $M$Te$_2$ as well as $M$P$_2$ ($M$=W, Mo) and the ternary variants of $MX$Te$_4$ ($M$ = Ta, Nb; $X$ = Ir, Rh) [5, 11, 14-16]. The electronic structure of the WSMs gives rise to fascinating phenomena in transport properties, such as a chiral anomaly in the presence of parallel electric and magnetic fields leading to negative magnetoresistance, a novel anomalous Hall response, surface-state quantum oscillations and exotic superconductivity [17-24]. These properties of the WSMs establish them as a new platform for exploring potential applications in nonlinear optics.

Similar to the lattice structure of TaIrTe$_4$, NbIrTe$_4$ also crystallizes in an orthorhombic (OR) unit cell with the *Pmn2$_1$* space group and can be viewed as a cell-doubling derivative [5, 15]. Theoretical calculations indicate that NbIrTe$_4$ has a more complicated electronic structure (16 Weyl points in the Brillouin zone) [5] than TaIrTe$_4$ (4 Weyl points in the Brillouin zone) [15, 25]. The presence of Weyl points in NbIrTe$_4$, accompanied by the common features of a large non-saturating *MR* effect, have been supported experimentally [26]. Pressure is an important tuning parameter that can be utilized to alter the lattice and corresponding electronic structures of a material and therefore influence its electrical transport properties. It is our goal to gain a deeper insight into the nature of Weyl semimetals and in particular to achieve a better understanding of the connection between its nontrivial WSM and SC states. Compelling examples of pressure-induced superconducting transition have been observed in the type-II WSM WTe$_2$ [27, 28], $T_d$-MoTe$_2$ and MoTe$_{2-x}$S$_x$, [20, 29, 30]. As a ternary variant of WTe$_2$, TaIrTe$_4$ has also displayed superconductivity at high pressures [31] and surface superconductivity [32]. The high-pressure behavior of the new WSM NbIrTe$_4$ remains unknown, and the connection between the WSM and SC states in the ternary WSMs as a function of lattice changes is an open question. In this work, we reveal the evolution of pressure-induced superconductivity from the ambient pressure WSM state in NdIrTe$_4$ and compare our results with those observed in TaIrTe$_4$.

High-quality single crystals of NbIrTe$_4$ were grown by a Te flux method as described in [26]. The high-pressure experimental details can be found in the Supplementary Material [33]. Figures 1a and 1b show the temperature ($T$) dependence of the electrical resistance ($R$) measured in a NbIrTe$_4$ sample (S#1) under pressures to 54.8 GPa within the temperature range of 1.5- 300 K. Over the pressure range investigated, $R(T)$ displays metallic behavior as observed at ambient pressure. To explore the potential superconducting transition in NbIrTe$_4$ at lower temperatures, we loaded a second sample (S#2) into the high-pressure cell and performed resistance measurements in a $^3$He cryostat down to 0.4 K. A resistance drop with a midpoint temperature of ~0.45 K is found at 27.2 GPa (Fig.1c). Upon further compression, a zero-resistance state is observed at pressures around 32.4 GPa (Fig. 1c), indicating a superconducting transition. The transition temperature increases with pressure and reaches 1.83 K at 54.6 GPa. To further characterize whether the pressure-induced resistance drop is associated with a superconducting transition, we applied a magnetic field along the inter planar direction at 57 GPa. As shown in Fig. 1d, the transition shifts to lower temperatures upon increasing the magnetic field, indicating that the observed resistance drop indeed originates from a superconducting transition. We estimated the upper critical magnetic field ($H_{c2}$) for the superconducting phase of NbIrTe$_4$ by using the Werthamer-Helfand-Hohenberg (WHH) formula [34]: $H_{C2}^{WHH}(0) = -0.693T_c(dH_{C2}/dT)_{T=T_c}$, which yields an upper critical field of about 1T at 57 GPa (inset of Fig. 1d).

To understand the connection between the superconductivity and the WSM state in NbIrTe$_4$, we performed high-pressure measurements of the Hall resistance ($R_{xy}$) by sweeping the magnetic field ($B$), which is applied perpendicular to the *ab*-plane, from 0 to 7 T at 10 K. As shown in Fig. 2a, $R_{xy}(B)$ is negative within the pressure range investigated, indicating that electrons are the dominant charge carriers at the Fermi surface. However, upon increasing pressure the slope of $R_{xy}(B)$ becomes smaller and approaches zero at pressures above 40 GPa. These results suggest that the shrinkage of the lattice by applying pressure likely adjusts the carrier density in these two materials, leading to $R_{xy}(B)=0$ eventually.

The emergence of superconductivity in WSMs is closely associated with a suppression of *MR* [27, 28, 31]. To reveal the connection between superconductivity and positive *MR* effect in NbIrTe$_4$, we carried out *MR* measurements on our sample at 10 K in the pressure range of 1.2 - 54.8 GPa. As shown in Fig.2b, the ambient-pressure NbIrTe4 shows a positive *MR* (*MR* (%) = 35% at 1.2 GPa, here *MR* (%) is determined by [*R(7T)-R(0T)*]/*R(0T)*]×100%) which can be suppressed dramatically by pressure. In the presence of superconductivity at ~27 GPa, the value of *MR* (%) is about 3.3 %.

The transport property of materials is usually influenced by the type of crystal structure. To understand why the value of *MR* does not reach zero at the critical pressure of superconducting transition, we performed high-pressure X-ray diffraction measurements. As shown in Fig. 3, all diffraction peaks gradually shift to higher angles upon increasing pressure, and the patterns can be indexed well with the orthorhombic structure (*Pmn2$_1$* space group) below 24.8 GPa. However, there are two new peaks present at 27.5 GPa (Fig. 2b, as indicated by stars). The intensity of these new peaks becomes more pronounced with elevating pressure. Because only two peaks were detected, we are not able to determine the structure for this new phase. As a result, we define the new phase as the high-pressure (HP) phase in this study. These results show that the HP phase emerged from the matrix of the ambient-pressure (OR) phase and coexisted with the OR phase in the pressure range of 27.5- 40.4 GPa is responsible for the non-zero MR state and presence of superconducting state. When pressure is increased above 40.4 GPa, we cannot index the peaks by the OR phase, implying that the OR phase completely transforms to the HP phase at this pressure, where MR reaches zero. In the sister compound TaIrTe$_4$, the pressure-induced superconducting state is also accompanied by a lattice distortion [31]. These results suggest that the structural change from the OR phase to the HP phase are responsible for the superconductivity in NbIrTe$_4$. In addition, it may be of great interest to investigate the transport properties and the electronic state at the boundary of OR phase and HP phase, where the WSM and SC states coexist.

We summarize our high-pressure experimental results in Fig.4. As shown in Fig.4a, three distinct regimes can be seen in the *P-T* diagram: the Weyl semimetal (WSM) state (on the left), the mixed state of the WSM and SC states (in the middle), and the SC state (on the right). In the left regime (below the critical pressure ($P_{c1}$), NbIrTe$_4$ remains in the OR structure (Fig.3) and its ground state is a WSM. Within this pressure range, $R_H$ displays a slow variation (Fig.4a) but *MR* shows a monotonic decrease with increasing pressure (Fig. 4 c). At $P_{c1}$, a HP phase emerges from the matrix of the OR phase, accompanied by superconductivity. Concomitantly, the value of $R_H$ starts to approach zero, and *MR* (%) decreases to 3.3 % (Fig.4b and 4c). In the middle of regime (27-40 GPa), a mixed state composed of the WSM state, characterized by the OR phase (Fig.3) as well as positive *MR* (Fig.2b), and the SC state is observed. In the high-pressure regime (40-57 GPa), *MR* (%) of NbIrTe$_4$ reaches zero and the OR-to-HP phase transition finishes (Fig.2b and Fig.3), implying that the ground state becomes a pure SC state.

To visualize the high-pressure behavior of these two materials, we compare the data obtained from NbIrTe$_4$ and TaIrTe$_4$ (see black squares taken from Ref. 31) in Fig.4. The critical pressure ($P_{c1}$) of the superconducting transition for NbIrTe$_4$ and TaIrTe$_4$ is almost same. Although the sign of their ambient-pressure $R_H$ is different, both of the $R_H$ change their trend when superconductivity appears. Previous theoretical calculations on TaIrTe$_4$ demonstrated that the topological band structure is remarkably modified by a volume shrinkage [25]. We therefore propose that NbIrTe$_4$ should share a similar mechanism to that of TaIrTe$_4$ because NbIrTe$_4$ shows a similar high-pressure behavior to that of TaIrTe$_4$. Moreover, at the critical pressure ($P_{c2}$) and above, not only MR (%) of the two materials reaches zero, but also their $R_H$ and $T_c$ surprisingly merge into the same values. These results suggest that, although NbIrTe$_4$ and TaIrTe$_4$ possess dramatically different band structures at ambient pressure, Nb and Ta in the ternary compound make the same contribution to the normal and SC states at high pressure. This is the first experimental case observed, showing the unconventional response of $T_c$ to remarkably different atomic mass ($m_{Nb}$=93, while $m_{Ta}$=181) but similar ion diameter. This fresh information may offer a potential method to "simulate the isotopic effect" for compounds with the different ion mass but the same size through application of pressure. Subsequently, a fundamental question is raised: what is the key factor for determining the superconducting and normal states in these pressurized WSMs, and does the superconductivity found in $M$IrTe$_4$ ($M$=Nb and Ta) can fall into the unconventional superconductivity category, which deserve further investigations from theoretical and experimental sides.

In summary, we report the first observation of superconductivity in pressurized type-II WSM candidate NbIrTe$_4$, through the complementary measurements of high-pressure resistance, Hall coefficient, magnetoresistance and synchrotron X-ray diffraction. The pressure-induced superconductivity state with the transition temperature ($T_c$) of 0.45 K emerges from the matrix of WSM state at 27.2 GPa. The WSM and superconducting (SC) states coexist in the pressure range of 27-40 GPa, then a pure SC state remains in the range of 40- 57 GPa. By comparing NbIrTe$_4$ and TaIrTe$_4$, we find a remarkably similar evolution process, from a WSM-OR phase to a SC-HP phase, in these two compounds. Intriguingly, the values of $R_H$, MR (%) and $T_c$ of NbIrTe$_4$ and TaIrTe$_4$ seem the same at ~ 40 GPa and above. Our high-pressure results not only reveal the universal connection among the lattice structure, superconductivity and the band structure in the WSMs but also demonstrate for the first time that pressure can tune these two WSMs to have the nearly identical normal and superconducting property. Such an unconventional behavior of superconductivity, independence of the mass of Nb and Ta ions, discovered in this study is of great interest and deserves investigations in the future.

**Acknowledgements**

The work in China was supported by the National Key Research and Development Program of China (Grant No. 2017YFA0302900, 2016YFA0300300 and 2017YFA0303103), the NSF of China (Grants No. U2032214, 11888101 and 12004419) and the Strategic Priority Research Program (B) of the Chinese Academy of Sciences (Grant No. XDB25000000). We thank the support from the Users with Excellence Program of Hefei Science Center CAS (2020HSC-UE015). Part of the work is supported by the Synergic Extreme Condition User System. S.C. and J. G. are grateful for support from the China Postdoctoral Science Foundation (E0BK111) and the Youth Innovation Promotion Association of the CAS (2019008). L.B. is supported by DOE-BES through award DE-SC0002613. The work at Los Alamos National Laboratory was supported by the Center for Integrated Nanotechnologies, and Office of Science User Facility operated for the U.S. Department of Energy, Office of Science. A portion of this work was performed at the National High Magnetic Field Laboratory, which is supported by the NSF Cooperative Agreement No. DMR-1644779, the U.S. DOE and the State of Florida.

* These authors contributed equally to this work.
†Corresponding authors
llsun@iphy.ac.cn

## References

[1] X. Wan, A. M. Turner, A. Vishwanath and S. Y. Savrasov. *Topological semimetal and Fermi-arc surface states in the electronic structure of pyrochlore iridates*. Physical Review B **83**, 205101 (2011).

[2] A. A. Burkov and L. Balents. *Weyl Semimetal in a Topological Insulator Multilayer*. Physical Review Letters **107**, 127205 (2011).

[3] S.-Y. Xu, I. Belopolski, N. Alidoust, M. Neupane, G. Bian, C. Zhang, R. Sankar, G. Chang, Z. Yuan, C.-C. Lee, S.-M. Huang, H. Zheng, J. Ma, D. S. Sanchez, B. Wang, A. Bansil, F. Chou, P. P. Shibayev, H. Lin, S. Jia and M. Z. Hasan. *Discovery of a Weyl*

*fermion semimetal and topological Fermi arcs*. Science **349**, 613 (2015).

[4] L. Lu, Z. Wang, D. Ye, L. Ran, L. Fu, J. D. Joannopoulos and M. Soljačić. *Experimental observation of Weyl points*. Science **349**, 622 (2015).

[5] L. Li, H.-H. Xie, J.-S. Zhao, X.-X. Liu, J.-B. Deng, X.-R. Hu and X.-M. Tao. *Ternary Weyl semimetal NbIrTe$_4$ proposed from first-principles calculation*. Physical Review B **96**, 024106 (2017).

[6] M. Z. Hasan and C. L. Kane. *Colloquium: Topological insulators*. Reviews of Modern Physics **82**, 3045-3067 (2010).

[7] X.-L. Qi and S.-C. Zhang. *Topological insulators and superconductors*. Reviews of Modern Physics **83**, 1057-1110 (2011).

[8] H. Weng, C. Fang, Z. Fang, B. A. Bernevig and X. Dai. *Weyl Semimetal Phase in Noncentrosymmetric Transition-Metal Monophosphides*. Physical Review X **5**, 011029 (2015).

[9] G. Xu, H. Weng, Z. Wang, X. Dai and Z. Fang. *Chern Semimetal and the Quantized Anomalous Hall Effect in HgCr$_2$Se$_4$*. Physical Review Letters **107**, 186806 (2011).

[10] S.-M. Huang, S.-Y. Xu, I. Belopolski, C.-C. Lee, G. Chang, B. Wang, N. Alidoust, G. Bian, M. Neupane, C. Zhang, S. Jia, A. Bansil, H. Lin and M. Z. Hasan. *A Weyl Fermion semimetal with surface Fermi arcs in the transition metal monopnictide TaAs class*. Nature Communications **6**, 7373 (2015).

[11] A. A. Soluyanov, D. Gresch, Z. Wang, Q. Wu, M. Troyer, X. Dai and B. A. Bernevig. *Type-II Weyl semimetals*. Nature **527**, 495-498 (2015).

[12] L. X. Yang, Z. K. Liu, Y. Sun, H. Peng, H. F. Yang, T. Zhang, B. Zhou, Y. Zhang,

Y. F. Guo, M. Rahn, D. Prabhakaran, Z. Hussain, S. K. Mo, C. Felser, B. Yan and Y. L. Chen. *Weyl semimetal phase in the non-centrosymmetric compound TaAs*. Nature Physics **11**, 728-732 (2015).

[13] B. Q. Lv, H. M. Weng, B. B. Fu, X. P. Wang, H. Miao, J. Ma, P. Richard, X. C. Huang, L. X. Zhao, G. F. Chen, Z. Fang, X. Dai, T. Qian and H. Ding. *Experimental Discovery of Weyl Semimetal TaAs*. Physical Review X **5**, 031013 (2015).

[14] Y. Sun, S.-C. Wu, M. N. Ali, C. Felser and B. Yan. *Prediction of Weyl semimetal in orthorhombic $MoTe_2$*. Physical Review B **92**, 161107 (2015).

[15] K. Koepernik, D. Kasinathan, D. V. Efremov, S. Khim, S. Borisenko, B. Büchner and J. van den Brink. *$TaIrTe_4$: A ternary type-II Weyl semimetal*. Physical Review B **93**, 201101 (2016).

[16] G. Autès, D. Gresch, M. Troyer, A. A. Soluyanov and O. V. Yazyev. *Robust Type-II Weyl Semimetal Phase in Transition Metal Diphosphides $XP_2$ (X=Mo, W)*. Physical Review Letters **117**, 066402 (2016)

[17] X. Huang, L. Zhao, Y. Long, P. Wang, D. Chen, Z. Yang, H. Liang, M. Xue, H. Weng, Z. Fang, X. Dai and G. Chen. *Observation of the Chiral-Anomaly-Induced Negative Magnetoresistance in 3D Weyl Semimetal TaAs*. Physical Review X **5**, 031023 (2015).

[18] A. A. Zyuzin and A. A. Burkov. *Topological response in Weyl semimetals and the chiral anomaly*. Physical Review B **86**, 115133 (2012).

[19] A. C. Potter, I. Kimchi and A. Vishwanath. *Quantum oscillations from surface Fermi arcs in Weyl and Dirac semimetals*. Nature Communications **5**, 5161 (2014).

[20] Y. Qi, P. G. Naumov, M. N. Ali, C. R. Rajamathi, W. Schnelle, O. Barkalov, M. Hanfland, S.-C. Wu, C. Shekhar, Y. Sun, V. Süß, M. Schmidt, U. Schwarz, E. Pippel, P. Werner, R. Hillebrand, T. Förster, E. Kampert, S. Parkin, R. J. Cava, C. Felser, B. Yan and S. A. Medvedev. *Superconductivity in Weyl semimetal candidate MoTe$_2$*. Nature Communications **7**, 11038 (2016).

[21] F. C. Chen, X. Luo, R. C. Xiao, W. J. Lu, B. Zhang, H. X. Yang, J. Q. Li, Q. L. Pei, D. F. Shao, R. R. Zhang, L. S. Ling, C. Y. Xi, W. H. Song and Y. P. Sun. *Superconductivity enhancement in the S-doped Weyl semimetal candidate MoTe$_2$*. Applied Physics Letters **108**, 162601 (2016).

[22] L. Zhu, Q.-Y. Li, Y.-Y. Lv, S. Li, X.-Y. Zhu, Z.-Y. Jia, Y. B. Chen, J. Wen and S.-C. Li. *Superconductivity in Potassium-Intercalated Td-WTe$_2$*. Nano Letters **18**, 6585-6590 (2018).

[23] V. Fatemi, S. Wu, Y. Cao, L. Bretheau, Q. D. Gibson, K. Watanabe, T. Taniguchi, R. J. Cava and P. Jarillo-Herrero. *Electrically tunable low-density superconductivity in a monolayer topological insulator*. Science **362**, 926 (2018).

[24] E. Sajadi, T. Palomaki, Z. Fei, W. Zhao, P. Bement, C. Olsen, S. Luescher, X. Xu, J. A. Folk and D. H. Cobden. *Gate-induced superconductivity in a monolayer topological insulator*. Science **362**, 922 (2018).

[25] X. Zhou, Q. Liu, Q. Wu, T. Nummy, H. Li, J. Griffith, S. Parham, J. Waugh, E. Emmanouilidou, B. Shen, O. V. Yazyev, N. Ni and D. Dessau. *Coexistence of tunable Weyl points and topological nodal lines in ternary transition-metal telluride TaIrTe$_4$*.

Physical Review B **97**, 241102 (2018).

[26] R. Schönemann, Y.-C. Chiu, W. Zheng, V. L. Quito, S. Sur, G. T. McCandless, J. Y. Chan and L. Balicas. *Bulk Fermi surface of the Weyl type-II semimetallic candidate NbIrTe$_4$*. Physical Review B **99**, 195128 (2019).

[27] D. Kang, Y. Zhou, W. Yi, C. Yang, J. Guo, Y. Shi, S. Zhang, Z. Wang, C. Zhang, S. Jiang, A. Li, K. Yang, Q. Wu, G. Zhang, L. Sun and Z. Zhao. *Superconductivity emerging from a suppressed large magnetoresistant state in tungsten ditelluride*. Nature Communications **6**, 7804 (2015).

[28] X.-C. Pan, X. Chen, H. Liu, Y. Feng, Z. Wei, Y. Zhou, Z. Chi, L. Pi, F. Yen, F. Song, X. Wan, Z. Yang, B. Wang, G. Wang and Y. Zhang. *Pressure-driven dome-shaped superconductivity and electronic structural evolution in tungsten ditelluride*. Nature Communications **6**, 7805 (2015).

[29] Z. Guguchia, F. von Rohr, Z. Shermadini, A. T. Lee, S. Banerjee, A. R. Wieteska, C. A. Marianetti, B. A. Frandsen, H. Luetkens, Z. Gong, S. C. Cheung, C. Baines, A. Shengelaya, G. Taniashvili, A. N. Pasupathy, E. Morenzoni, S. J. L. Billinge, A. Amato, R. J. Cava, R. Khasanov and Y. J. Uemura. *Signatures of the topological s+− superconducting order parameter in the type-II Weyl semimetal Td-MoTe$_2$*. Nature Communications **8**, 1082 (2017).

[30] Y. Li, Q. Gu, C. Chen, J. Zhang, Q. Liu, X. Hu, J. Liu, Y. Liu, L. Ling, M. Tian, Y. Wang, N. Samarth, S. Li, T. Zhang, J. Feng and J. Wang. *Nontrivial superconductivity in topological MoTe$_{2-x}$S$_x$ crystals*. Proceedings of the National Academy of Sciences **115**, 9503 (2018).

[31] S. Cai, E. Emmanouilidou, J. Guo, X. Li, Y. Li, K. Yang, A. Li, Q. Wu, N. Ni and L. Sun. *Observation of superconductivity in the pressurized Weyl-semimetal candidate TaIrTe$_4$*. Physical Review B **99**, 020503 (2019).

[32] Y. Xing, Z. Shao, J. Ge, J. Luo, J. Wang, Z. Zhu, J. Liu, Y. Wang, Z. Zhao, J. Yan, D. Mandrus, B. Yan, X.-J. Liu, M. Pan and J. Wang. *Surface superconductivity in the type II Weyl semimetal TaIrTe4*. National Science Review **7**, 579-587 (2020).

[33] *Supplemental Material for the high pressure experimental details*.

[34] N. R. Werthamer, E. Helfand and P. C. Hohenberg. *Temperature and Purity Dependence of the Superconducting Critical Field, H$_{c2}$. III. Electron Spin and Spin-Orbit Effects*. Physical Review **147**, 295-302 (1966).

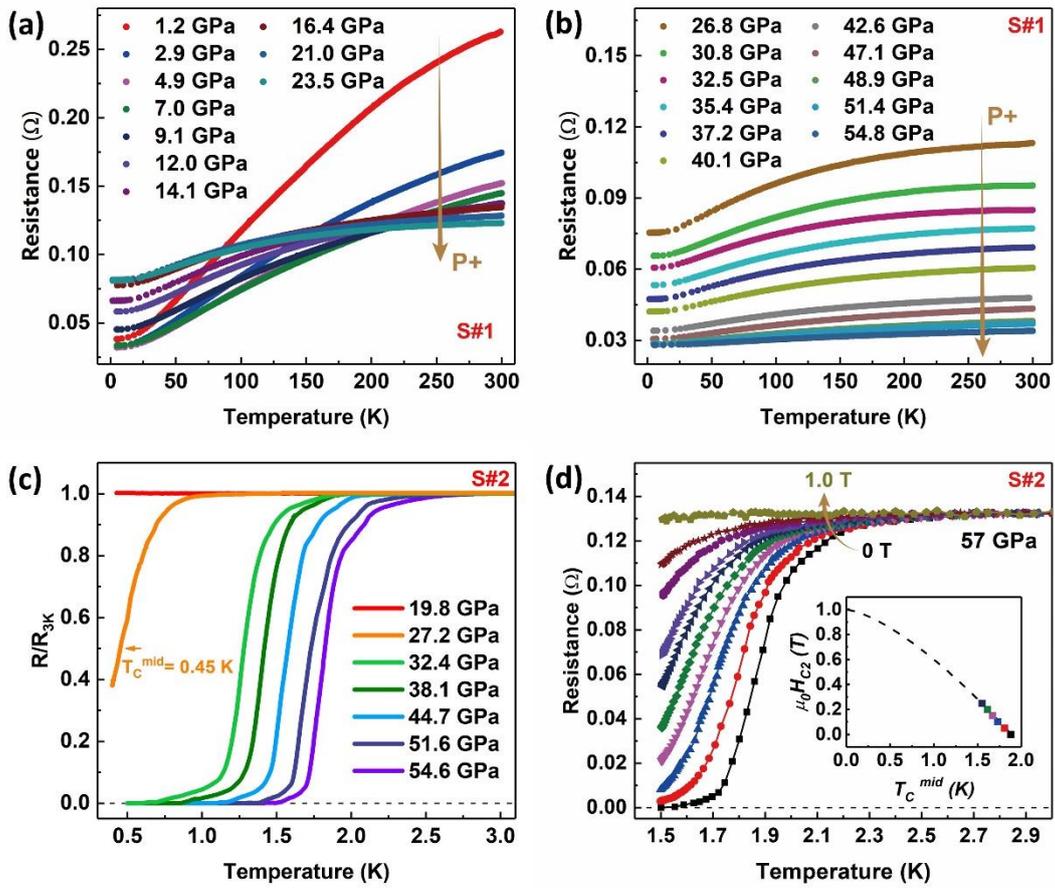

Figure 1. Temperature-dependence of electrical resistance at different pressures and magnetic fields. (a) and (b) The resistance as a function of temperature for the sample 1 (S#1) measured over the range of 1.5-300 K at different pressures. (c) The resistance versus temperature for the sample 2 (S#2) measured in the range of 0.3-3 K at different pressures, showing the superconducting transition. (d) Magnetic field dependence of the superconducting transition temperature measured at 57 GPa. The inset shows the dependence of upper critical field with midpoint superconducting transition temperature ($T_c^{mid}$).

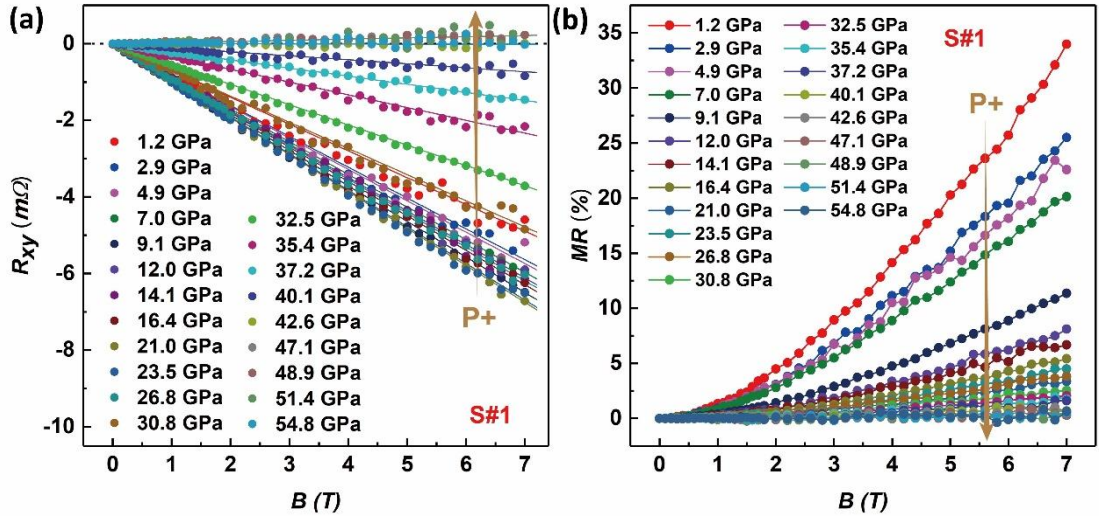

Figure 2. Results of Hall and magnetoresistance obtained from NbIrTe$_4$. (a) Hall resistance ($R_{xy}$) as a function of magnetic field at different pressures, displaying that $R_{xy}(B)$ exhibits a slow change below 26.8 GPa, then increases above this pressure and tends to zero upon further compression. (b) Magnetic field dependence of magnetoresistance ($MR(\%)$) for pressures ranging from 1.2 GPa to 54.8 GPa (here $MR(\%)=[R(7T)-R(0T)]/R(0T) \times 100\%$), showing that $MR(\%)$ decreases with increasing pressure.

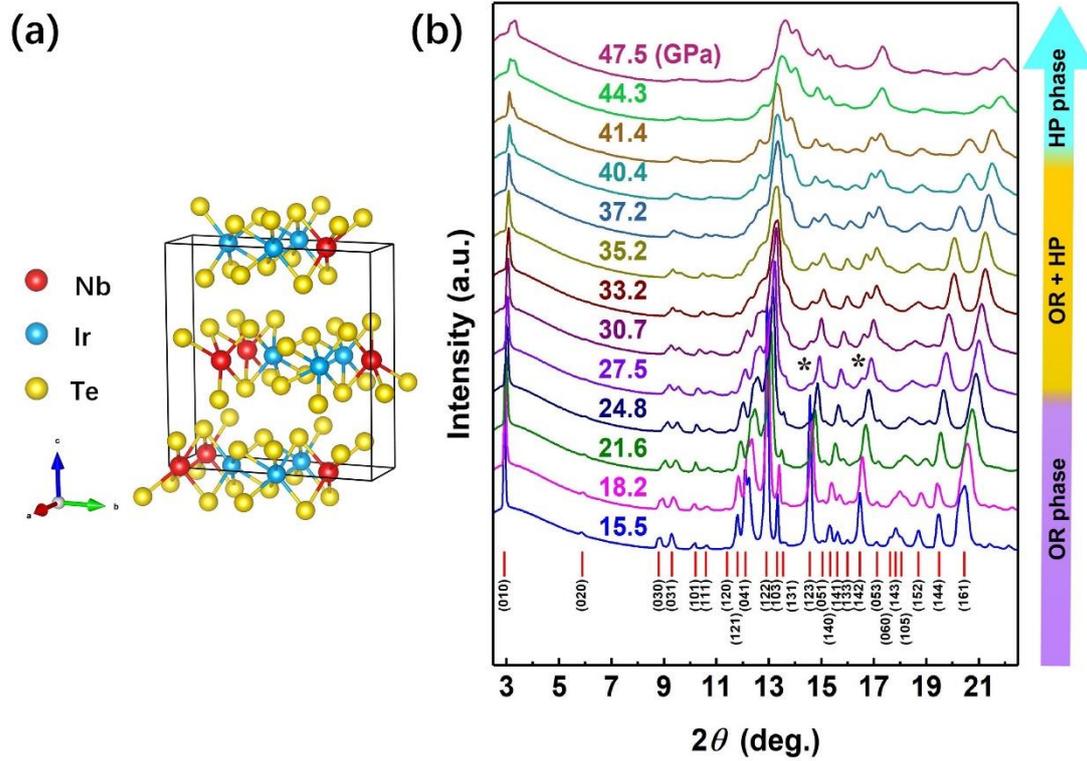

Figure 3. (a) Schematic crystal structure of NbIrTe$_4$. (b) X-ray diffraction patterns collected at different pressures. In the range of 15-27 GPa, the sample remains OR phase (see violet regime), while, in the range of 27.5-40.4 GPa, it hosts a mixed phase composed of OR and HP phases (see the orange regime). Upon further compression, the sample possesses a pure HP phase (see the cyan regime).

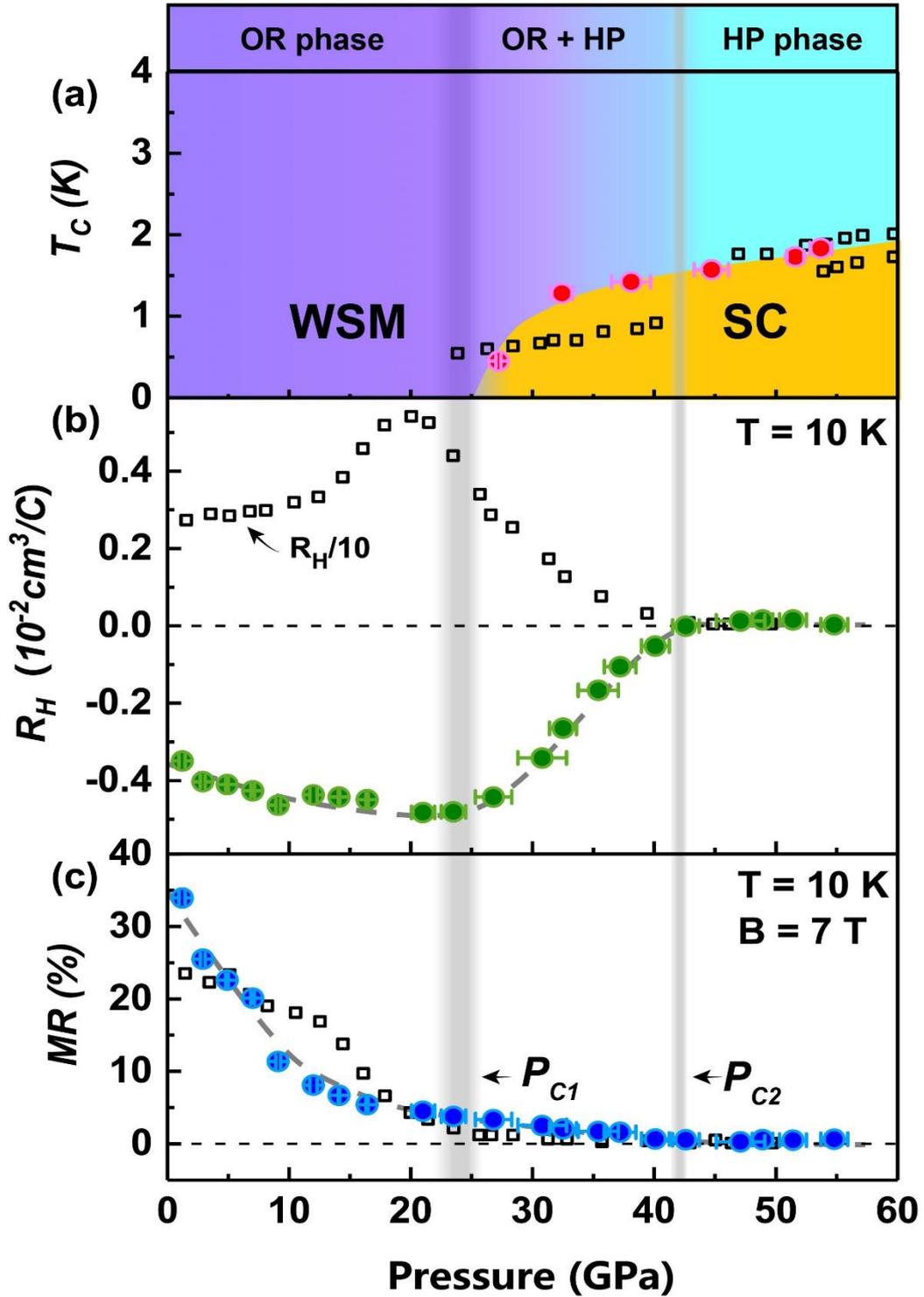

Figure 4. Summary of experimental results of NbIrTe$_4$ and comparison with TaIrTe$_4$. (a) Pressure −$T_c$ phase diagram with structure information. The data of TaIrTe$_4$ (see black squares taken from Ref. 31) is included. WSM and SC represent Weyl semimetal and superconducting states, respectively. (b) Pressure dependence of Hall coefficient ($R_H$)

measured at 10 K. (c) Magnetoresistance (*MR*) as a function of pressure measured at 10 K, where $MR(\%) = [R(7T)-R(0T)]/R(0T) \times 100\%$. The black squares in figure (b) and (c) are the data of TaIrTe$_4$ [31]).